\documentclass[letterpaper]{aa}

\input epsf.sty
\usepackage{graphicx}
\usepackage{txfonts}

\begin{document}
  
  \title{Log-parabolic spectra and particle acceleration in blazars. \\ 
    III: SSC emission in the TeV band from Mkn~501}
  \author{E.~Massaro\inst{1}
    \and A.~Tramacere\inst{1}
    \and M.~Perri\inst{2}
    \and P.~Giommi\inst{2} 
    \and G.~Tosti\inst{3} 
  }

  \offprints{enrico.massaro@uniroma1.it}

  \institute{
    Dipartimento di Fisica, Universit\`a La Sapienza, Piazzale A. Moro 2, 
    I-00185 Roma, Italy
    \and ASI Science Data Center, ESRIN, I-00044 Frascati, Italy
    \and Dipartimento di Fisica, Universit\`a di Perugia, via A. Pascoli, 
    Perugia, Italy 
  }

  \markboth{E. Massaro et al.: Log-parabolic spectra and particle acceleration in blazars. III: 
    SSC emission in the TeV band from Mkn~501 }
	   {E. Massaro et al.:  Log-parabolic spectra and particle acceleration in blazars. III: 
	     SSC emission in the TeV band from Mkn~501}
  
  \date{Received ... ; Accepted ...}
  
  \abstract{
Curved broad-band spectral distributions of non-thermal sources like
blazars are described well by a log-parabolic law where the second
degree term measures the curvature. 
Log-parabolic energy spectra can be obtained for relativistic electrons
by means of a statistical acceleration mechanism whose probability of 
acceleration depends on energy. 
In this paper we compute the spectra radiated by an electron population via synchrotron
and Synchro-Self Compton processes to derive the relations between 
the log-parabolic parameters. 
These spectra were obtained by means of an accurate numerical code 
that takes the proper spectral distributions for single particle 
emission into account. 
We found that the ratio between the curvature parameters of the synchrotron spectrum 
to that of the electrons is equal to $\sim$0.2 instead of 0.25, the value foreseen 
in the $\delta$-approximation.
Inverse Compton spectra are also intrinsically curved and can be approximated 
by a log-parabola only in limited ranges. 
The curvature parameter, estimated around the SED peak, may vary from a lower value than 
that of the synchrotron spectrum up to that of emitting electrons depending on 
whether the scattering is in the Thomson or in the Klein-Nishina regime.
We applied this analysis to computing the synchro-self Compton emission from the BL Lac 
object Mkn 501 during the large flare of April 1997. We fit simultaneous {\it Beppo}SAX 
and CAT data and reproduced intensities and spectral curvatures of both components 
with good accuracy. 
In particular, the large curvature observed in the TeV range was found to be mainly 
intrinsic, and therefore did not require a large pair production absorption against the 
extragalactic background. We regard this finding as an indication that the Universe is 
more transparent at these energies than previously assumed by several models found 
in the literature.
This conclusion is supported by recent detection of two relatively
high redshift blazars with H.E.S.S.
}
\authorrunning{E. Massaro et al.}
\titlerunning{Log-parabolic spectra in blazars III: SSC emission in 
the TeV band from Mkn~501
}
\maketitle

\section{Introduction}
The spectral energy distributions (SEDs) of blazars, and particularly of 
BL Lac objects, are characterised by a double-peak structure. 
The peak at lower energies is generally explained by synchrotron radiation 
from relativistic electrons in a jet that is closely aligned to the line of sight, 
while the high frequency bump is produced by inverse Compton scattering 
from the same electron population interacting either with the synchrotron 
photons (SSC, Synchrotron-Self Compton models, see e.g. Marscher  \& Gear, 
1985) or with other photons that originated in the local environment 
(ERC, External Radiation Compton models, see e.g. Sikora, Begelman \& 
Rees, 1994). 

For the so-called Low-energy peaked BL Lac (LBL) objects, the frequency 
of the first peak is in the Infrared-Optical region, while it is in 
the UV-X ray range for the High-energy peaked BL Lac (HBL) (Padovani and 
Giommi 1995).
The shape of these bumps is characterised in the $Log(\nu F_{\nu})$ vs. 
$Log~\nu$ plot by a rather smooth curvature extending through several 
frequency decades. 
A very simple and successful analytical function that can model the shape 
of these broad peaks is a parabola in the logarithms of the variables 
(hereafter log-parabola). 
This function has only three spectral parameters, and was used by Landau et 
al. (1986) to fit the broad band spectra of some bright BL Lac objects from 
radio to UV. 
Sambruna, Maraschi \& Urry (1996) also used a log-parabola to fit
blazars' SED to estimate peak frequencies and luminosities without, however, 
attributing a physical meaning to this shape.
In two previous papers (Massaro et al. 2004a, 2004b, hereafter Paper I 
and Paper II, respectively) we used the log-parabolic model to fit the 
BeppoSAX broad band X-ray spectra of the BL Lac objects Mkn~421 and 
Mkn~501 and studied the relations between the spectral parameters and the
luminosity. 
We also showed that, under simple approximations, a log-parabolic 
synchrotron spectrum can be obtained by a relativistic electron population
having a similar energy distribution, and then derived the main relations 
between their parameters.
Furthermore, in Paper I we proposed a simple explanation for the log-parabolic 
energy distribution of the electrons as resulting from a statistical 
mechanism where the acceleration probability decreases with the particle 
energy.

Curved spectra in non-thermal sources have already been studied in the past
and generally were related to the radiative ageing of the emitting electrons
that have a single power law injection spectrum. 
The well-known solution by Kardashev (1962) predicts a change in the 
electron spectral index by a unity (0.5 for the synchrotron
spectrum) around a break energy, whose value decreases with time.
Probably for this reason, curved spectra have been modelled by means of
rather complex functions like a double broken power law (Kataoka et al. 1999)
or a continuous combination of two power laws after releasing the condition
on the difference of spectral indices (see the discussion in Fossati et
al. 2000). 
A similar spectrum is generally used in the spectral modelling
of gamma-ray bursts (Band et al. 1993). 
On the same basis, Sohn et al. (2003) more recently introduced a 
spectral-curvature parameter, defined as the difference between two
spectral indices in different frequency intervals, to study the evolution 
of synchrotron sources.
As we will show later, this parameter is strongly dependent on the chosen 
frequency range and cannot be univocally related to the actual curved
shapes.

In this paper we stress the advantages of using a log-parabolic law 
to model curved spectra over broad frequency ranges and present the 
results of an accurate study of the spectral distributions of the 
synchrotron radiation (hereafter SR) and of the inverse Compton (hereafter 
IC) scattered photons by a population of relativistic electrons with 
such energy spectrum. 
We study, in particular, the relations between the parameters of SR spectra 
and those of the electrons and show that they are generally 
approximated well by simple power behaviours and that they provide useful 
information on the electron spectrum. 
The relations with the IC spectrum are not so simple depending upon whether the
majority of scatterings is in the Thomson or in the Klein-Nishina regime.
In any case, our calculations indicate a good way to properly model
the observed spectra.
We also consider the case in which the probability of statistical 
acceleration is constant below a critical energy, thus producing a power 
law spectrum that turns into a log-parabola above this energy.

In the last part of the paper, we apply our model to the large flare 
from Mkn~501 observed in April 1997. 
This is a unique event in which the SR and IC components were 
observed simultaneously over energy bands that were broad enough to 
study the spectral curvature. 
We use the BeppoSAX data already analysed in 
Paper II together with those in the TeV band obtained by the CAT
experiment (Djannati-Atai et al. 1999). Our approach of studying
both the SR and IC spectral curvatures is useful for evaluating the
relevance of the intergalactic absorption due to pair production
from photon-photon interactions. In particular, we discuss the
possibility that the observed TeV curvature is mainly intrinsic, as
it implies rather low absorption and the possibility of detecting
sources at larger distances than usually estimated.  

\section{Properties of the log-parabolic spectral distribution}

The log-parabolic model is one of the simplest ways to represent curved 
spectra when these show mild and nearly symmetric curvature around the 
maximum instead of a sharp high-energy cut-off like that of an exponential. 
In the following, we summarise the main properties of log-parabolic 
spectra, while for details we refer to 
Papers I and II. This law has only one more parameter than a simple power law 
and can be written as:
\begin{equation}
 F(E) = K (E/E_1)^{-(a~+~b~ Log(E/E_1))} ~~~~~~~~  
\end{equation}
The energy dependent photon index is
\begin{equation}
 \alpha(E) = a~ +~ 2~ b~ Log(E/E_1) ~~~~~~~~.  
\end{equation}
With $\nu_p=E_p/h$ we also indicate the peak frequency 
of the SED where $E_p$ is related to the spectral parameters 
of Eq.~(1) as follows:
\begin{equation}
 E_p = E_1~ 10^{(2-a)/2b} ~~~~~~~~ 
\end{equation}
and 
\begin{equation}
\nu_p F(\nu_p) = K E_1 E_p (E_p/E_1)^{-a/2} = K E_1^2 ~10^{(2-a)^2/4b} ~~.
\end{equation}

It is useful to note that the above formulae must be used
when working with photon spectra, as is usual in X and $\gamma$-ray
astronomy. 
Energy spectra are obtained by multiplying Eq.(1) by $E$, and the 
corresponding spectral index is therefore increased by unity. 
In the following, we indicate the energy index with $a_e=a - 1$ to avoid confusion.
 In this case the peak frequency becomes:
\begin{eqnarray}
\nu_p = \nu_1~ 10^{(1-a_e)/2b} ~~~~~~~~~~~~~~~~~~~~~~~~~~~~~~~~~~~~~~~~~~~~~~ (3') . \nonumber 
\end{eqnarray}

An advantage of the log-parabolic model compared to other curved 
spectral laws is that the curvature around the peak is characterised only 
by the parameter $b$, while in other models, like the continuous combination 
of two power laws, it is a function of more parameters. 
Note also that the estimate of $b$ is independent of the particular spectral 
representation adopted for the data. 
The same property does not hold for the spectral-curvature parameter
(SCP) defined by Sohn et al. (2003). 
Unlike $b$, this quantity depends upon the energy 
(frequency) interval chosen to estimate the spectral indices at high 
$\alpha_h$ and low $\alpha_l$ frequencies. 
The relation between $b$ and SCP can be obtained by evaluating $\alpha$ in
two energy ranges $E_h$ (high) and $E_l$ (low) from Eq.(2):
\begin{equation}
{\rm SCP} = \frac{\alpha_h - \alpha_l}{\alpha_h + \alpha_l} = \frac{b~Log(E_h/E_l)}{a + b~Log(E_hE_l/E_1^2)} ~~~~~.
\end{equation}
For flat spectrum sources with $a$ close to zero, SCP becomes 
practically independent of $b$.
  
One limit of the log-parabolic model, however, is that it can represent 
only distributions symmetrically decreasing with respect to the 
peak frequency. 
However, it is not difficult to modify Eq.(1) to take into account a possible 
asymmetry with respect to $E_p$. 
For instance, one can use two different values of $b$ for energies lower 
and higher than $E_p$, where the two branches continuously match.  
An interesting possibility, which can be simply explained in terms of 
the energy dependence of the particle acceleration probability (see next 
section), is that the low energy segment of the spectrum follows a single 
power law with photon index $\alpha_0$ and that the log-parabolic 
bending becomes apparent only above a critical value $E_c$. 
This behaviour can be described by the following model that takes into 
account the continuity conditions on the flux and on $\alpha(E)$ at $E_c$: 
\begin{eqnarray}
F(E)&=&K(E/E_1)^{-\alpha_0}~~~,~~~~~~~~~ E\leq E_c \nonumber \\
F(E)&=&K(E/E_c)^{b~Log(E_c/E_1)}~\times \nonumber\\
 & & \times~(E/E_1)^{-(\alpha_0+b~Log(E/E_c))}~~, E > E_c ~~. 
\end{eqnarray}

\section{Statistical particle acceleration and log-parabolic spectra}
In Paper I we showed that a log-parabolic distribution is not only 
a simple mathematical tool for spectral modelling, but that it also relates to the 
physics of the statistical acceleration process under some simple hypotheses. 
In the following, we extend the considerations of Paper I. 
The results, useful to derive some relations between the acceleration 
parameters, should not be considered as a complete theory of energy-dependent
acceleration. 
To confirm the main findings of this approach, more detailed analytical 
and numerical calculations are necessary, but such a detailed theoretical 
approach is beyond the purpose of the present paper.

\subsection{Energy distribution of accelerated particles}

The energy spectrum of particles accelerated by some statistical mechanism, 
e.g. a shock wave or a strong perturbation moving down a jet, is given by 
a power law (Bell 1978, Blandford \& Ostriker 1978, Michel 1981).
In Paper I we showed that a log-parabolic 
energy spectrum is obtained when the condition that the acceleration probability
$p$ is independent of energy is released and that its value
at the step $i$ satisfies a power law relation as:
\begin{equation}
p_i = g/\gamma_i^q ~~~, ~~ (i=0,1,2,...)
\end{equation}
where $g$ and $q$ are positive constants.
Such a situation can occur, for instance, when particles are confined 
by a magnetic field with a confinement efficiency that decreases for an 
increasing gyration radius. 
In Paper I (Sect. 6.1) we found that the integral energy distribution of 
accelerated particles is given by a log-parabolic law:
\begin{equation}
N(>\gamma) = N_0 (\gamma/\gamma_0)^{-[s - 1 + r Log (\gamma/\gamma_0)]} ~~~~~~~,
\end{equation}
with
\begin{equation} 
s = - \frac{Log(g/\gamma_0^q)}{Log~\varepsilon} - \frac{q-2}{2} ~~~~~~~,
\end{equation}
\begin{equation}
r = \frac{q}{2~Log~\varepsilon} ~~~~~~ ,
\end{equation}
where $\gamma_0$ is the minimum Lorentz factor in Eq.(7).
As already discussed in Paper I the differential energy distribution 
$N(\gamma)$ is not an exact log-parabolic law, but differs from it 
only by a logarithmic term and can be approximated well by
a log-parabola with a very similar curvature over an energy range of
several decades. 

\begin{figure}
      \vspace{0.0cm}
      \hspace{0.5cm}
\epsfysize=8cm
\epsfbox{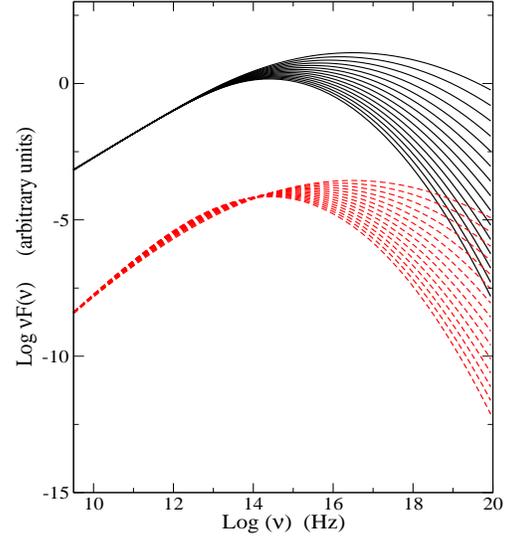}
\caption[]{
Synchrotron spectra emitted by an electron population
with log-parabolic (dashed lines) and power-law log-parabolic
energy distributions (Eq. 11) (solid lines), computed for $r$ 
values in the interval 0.50--1.20. Spectra were shifted
on the vertical scale to avoid confusion.
}
\label{fig1}
\end{figure}

Shock wave acceleration is not the only statistical mechanism active in 
non-thermal sources. 
For instance, electron acceleration can occur in magnetohydrodynamical
turbulence in which regions of magnetic field reconnection can develop 
in a very stochastic way. 
Recently, Nodes et al. (2004) presented the results of the numerical
simulations of a relativistic particle acceleration in a three-dimensional
turbulent electromagnetic field configuration, also taking their SR into account. 
These authors found energy spectral distributions that where significantly 
flatter than $s=2$ and, in a few cases, characterised by a steepening 
spectral index at high energies.
We verified that over sufficiently wide energy ranges, the spectra given by 
Nodes et al. (2004) are represented well by a log-parabolic law or by
a combination of a power law and a parabola. The resulting curvature
parameters are generally small, but probably the curvature depends
on the distribution and size of the acceleration regions, so one 
can expect that, under different assumptions, it could be higher. 
Nodes et al. (2004) computed the emerging SR spectra, which show an appreciable curvature 
over a few decade frequency range. 

Statistical acceleration is not the only way to obtain curved electron 
spectra. Energy distributions showing a rather mild
curvature have also been obtained in blazar physics. 
To model the SED of MeV blazars, Sikora et al. (2002) assumed that 
electrons are accelerated via a two-step process with a broken 
power-law energy distribution as injection.
When the cooling effects are taken into account, the resulting electron
spectrum (see, for instance, Fig. 6 in their paper) can be described well
by a log-parabola over a range that is wider than three decades,
as we verified (in this case we found $r$=0.44).    

The assumption of Eq.~(7) about the energy dependence of the 
acceleration probability can be modified to take other 
escape processes into account. 
For instance, one can assume that the acceleration probability is constant 
for low energies and that it begins to decrease above a critical Lorentz factor 
$\gamma_a$ (see, for instance Eq. (23) in Paper I).
An approximate expression for the energy distribution of accelerated 
particles under this condition 
for $\gamma \ll \gamma_a$ should follow a power law with spectral 
index $s_0 = -(Log~g)/(Log~\varepsilon)$, while for $\gamma \gg \gamma_a$ 
it will approximate a log-parabolic spectrum like Eq.(8):
\begin{eqnarray}
 N(\gamma)&=&N_0~(\gamma/\gamma_0)^{-s_0} ~~~~ \gamma\leq \gamma_0 \nonumber  \\
 N(\gamma)&=&N_0~(\gamma/\gamma_0)^{-(s_0+r~Log(\gamma/\gamma_0))} ~~ \gamma > \gamma_0 ~~~, 
\end{eqnarray}
where for the sake of simplicity we take $\gamma_0$=$\gamma_a$.
In addition to the power law and log-parabolic segments, we can also expect
a sharp high-energy cutoff that appears when losses make the acceleration 
process highly inefficient.

\section{ Spectral properties of the emitted radiation}
\subsection{Synchrotron spectrum}

It is important to know the relations between the spectral parameters 
$a$ and $b$ of the emitted radiation and those of the electron population, 
namely $s$ and $r$.  
The spectral distribution 
of the SR by relativistic electrons having a log-parabolic energy 
distribution cannot be computed analitically.
For our purpose, however, the relations between the spectral parameters
can be derived under the usual $\delta$--approximation and the 
assumption that the electrons are isotropically distributed in a 
homogeneous randomly oriented magnetic field with an average 
intensity $B$ 
\begin{equation}
j_S(\nu)=\int P(\nu(\gamma)) N(\gamma)~ d\gamma 
\end{equation}
where $j_S$ is the synchrotron emissivity, $N(\gamma)$ the electron
density, and the power radiated by a single particle is
\begin{equation}
P(\nu)= {\frac{4e^4}{9m^2c^3}}~ \gamma^2 B^2  ~\delta(\nu - \nu_S)~~~, 
\end{equation}
and
\begin{equation}
\nu_S~=~ 0.29 {\frac{3e}{4 \pi mc}}~  \gamma^2 B  
\end{equation}
is the SR peak frequency for a single particle.
We then obtain (see Paper I):
\begin{equation}
P_S(\nu) \propto N_0 B^2 (\nu/\nu_0)^{-(a + b Log(\nu/\nu_0))}    
\end{equation}
with
\begin{eqnarray}
 a~&=&~(s-1)/2  \nonumber \\  
 b~&=&~r/4 ~~~~~~.
\end{eqnarray}

Note that the parameter $a$, as defined above, 
does not coincide with the one defined in Eq.~(2), because the 
former is the energy index at the frequency $\nu_0$, whereas the 
latter is the photon index at the energy $E_1$, not necessarily
corresponding to $\nu_0$.
It is not difficult to verify that these two parameters differ for an 
additional constant, whose value depends on $\nu_0$ and $\nu_1=E_1/h$. 
In Paper I we noted that 
one can expect $a$ and $b$ to show a linear correlation, 
which was indeed found for Mkn~421. 
We stress that such correlation is not a general
result that can be applied to any sample of sources. 
It also depends upon the 
values of $q$, $g$, and $\gamma_0$, which can be different from source
to source, and the dispersions of their distribution functions can introduce
a significant amount of scatter in the data that could destroy the correlation. 
Reasonably, one has to expect that the linear correlation between 
$a$ and $b$ can be found only when the physical conditions in the 
acceleration region are quite similar. 
The absence of correlation found by Perlman et al. (2005) in their 
sample, therefore,  does not imply that curved spectra in blazars
are not originated by energy-dependent statistical acceleration.

\begin{figure}
      \vspace{0.0cm}
      \hspace{0.0cm}
\epsfysize=8cm
\epsfbox{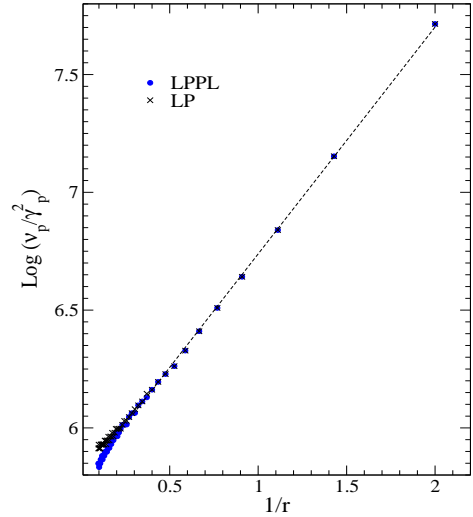}
\caption[]{
The relation between the peak frequencies of the synchrotron
SED and those of the energy spectrum of the emitting electrons
computed using the SFD (Eq. 17) for the LP and LPPL cases (see text). 
The abscissa is the inverse of the electron spectrum curvature 
to show the linear trend expected from Eq.(18). 
The dashed line is the best fit computed for $1/r >$0.33.
Note the small deviations of LPPL points from the best fit line
at low values of $1/r$.  
}
\label{fig2}
\end{figure}

The relation between $r$ and $b$ given above (Eq.16) is not exact:  
one can expect that a precise calculation of the spectral curvature 
of the emitted radiation must give a $b$ value smaller than in the 
$\delta$ approximation and that, for high values of $r$, $b$ can be 
greater than unity (depending on the frequency interval used around 
the peak in which it is estimated) because of the exponential cutoff 
by the SR spectrum radiated by a single particle.

To compute the relations between the spectral parameters numerically 
and to study the spectral evolution of the emitting particles, 
as reported in the next subsection, we applied a time-dependent 
numerical code (Tramacere \& Tosti, 2003; see
Tramacere, 2002, for a detailed description of the code). 
This code uses a numerical 
integration routine that applies a modified Simpson rule ensuring 
good convergence in short computing times.
In particular, this code includes an accurate evaluation of the synchrotron
frequency distribution (hereafter SFD):
\begin{equation}
P(\nu) =  \frac{\sqrt{3}e^3}{mc^2} B \frac{\nu}{\nu_c} \int_{\nu/\nu_c}^{\infty} K_{5/3}(x)~dx 
\end{equation}
where $K_{5/3}(x)$ is a modified Bessel function of the second type 
and $\nu_c$ the synchrotron critical frequency (Blumenthal
and Gould 1970, Rybicki \& Lightman 1979). 
Note that while the maximum of $P(\nu)$ is at 0.29 $\nu_c$, the peak in the corresponding 
SED $\nu P(\nu)$ is at the frequency $\nu_{pS}$ =1.33 $\nu_c$. 

\begin{figure}
      \vspace{0.0cm}
      \hspace{0.0cm}
\epsfysize=8cm
\epsfbox{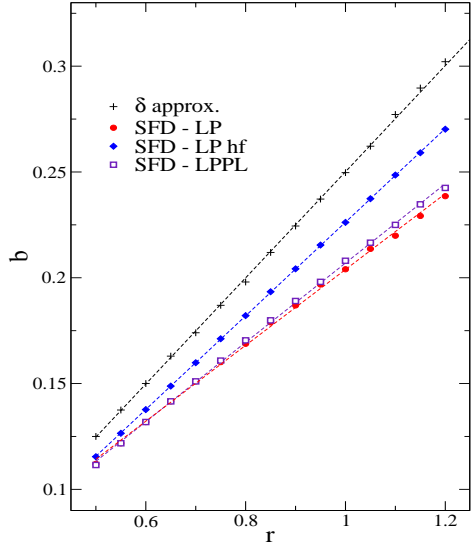}
\caption[]{
The relation between the log-parabolic curvature parameters
of the electron energy distribution $r$ and those of the 
synchrotron radiation computed applying the $\delta$ approximation 
(crosses) and the SFD. The two LP point sets correspond to 
log-parabolic electron spectra with $b$ evaluated by a best 
fit around the SED peak (filled circle) and on the high
frequency branch (diamonds). LPPL points (open squares)
correspond to a combined log-parabolic power law distribution.
Dashed lines are linear best fits used for the calculation of
the $b/r$ ratio.     
}
\label{fig3}
\end{figure}

We calculated the SR spectra and SEDs from Eq.(12) using a log-parabolic 
distribution without (LP) and with a low-energy power law branch (LPPL) 
(Eq. 11) and the frequency distribution of Eq.(17).
All the calculations in this and in the following subsections are 
performed in the co-moving frame, so relativistic beaming
effects were not included. 
 
Figure 1 shows some examples of the resulting SEDs.
In these computations we assumed a uniform magnetic field $B$=0.14 G, 
and the spectral parameters for the electrons were $\gamma_0$=10$^3$, 
$s$=$s_0$=1.2 (Eq. 11), while $r$ values were increased from 0.5 to 1.20 by 
steps of 0.05.
The electron density and the size of the emission region were taken
to be equal to the typical values used in blazar modelling.
In these conditions the synchrotron self-absorption is important
at frequencies lower than $\sim$10$^{10}$ Hz.
Note in Fig. 1 that at low frequencies both spectral sets tend to a 
power law: in the LP case the spectral index corresponds to the SFD 
asymptotic behaviour ($\nu^{1/3}$), while in the other case it has 
the energy spectral index $\alpha_{e0}=(s_0-1)/2$.
At high frequencies the log parabolic behaviour is very evident with a
peak frequency decreasing for higher values of $r$, as expected from 
Eq.(3) since $b$ is also increasing with $r$.   

We first verified that the peak frequency in the SED of the SR component
$\nu_{pS}$ is proportional to the Lorentz factor $\gamma_p$ at which 
the electron distribution $\gamma^2 ~N(\gamma)$ has its maximum. 
Applying the same formulae of Sect. 2, we have 
\begin{eqnarray}
\gamma_p = \gamma_0 10^{(2-s_0)/2r}    \nonumber
\end{eqnarray}
from which we have:
\begin{equation}
\frac{\nu_{pS}}{\gamma_p^2} = \frac{\nu_0}{\gamma_0^2}~ 10^{1/r} 
\end{equation}
where we assumed $\nu_0 = \nu_1$ for the sake of simplicity.
In Fig. 2 we plotted the resulting $\nu_{pS}/\gamma_p^2$ vs $1/r$
for both LP and LPPL cases: points of the two sets are coincident 
in the whole range of interest and only for $1/r <$ 0.3 a small
deviation from the linear trend of the LPPL model is evident.
The slope of the best fit line (for $1/r >$ 0.3) is equal to
0.97, very close to unity as in Eq.(18). 
Note also that the linear extrapolation to large values of
$r$ ($1/r \rightarrow 0$) gives 
for the $\nu_{pS}/\gamma_p^2$ ratio a value very close to 
1.33~$\nu_c$, consistent with the theory expectation.

We also used the spectra of Fig. 1 to study the relation between 
$r$ and $b$ under different conditions. 
The estimate of $b$ is not simple because the radiation spectra 
do not have an exact log-parabolic shape. 
One can expect that this estimate is more accurate considering only 
the SED branches at frequencies higher than the peak: a quadratic best 
fit of the spectra of Fig. 1 confirms the linear relation between $b$ 
and $r$, but the coefficient is equal to 0.22, slightly lower than 
the value found under the $\delta$ approximation (Eq. 16).
We also computed $b$ considering a frequency interval centred at the
peak frequency and spanning  approximately two decades. 
The results are shown in Fig. 3.
The relation between $b$ and $r$ was linear with very good accuracy, 
but the coefficient decreased to 0.18, as expected because of the 
smaller curvature of the low frequency branches.
The last result does not change if the low energy electron spectrum 
follows a power law instead of a log-parabola (open squares in Fig. 3): 
the $b/r$ ratio in this case was found equal to 0.19.
We also verified our results by computing the spectra using the $\delta$ 
approximation and found $b/r$=0.2508, in very good agreement with the 
theory.
An approximate rule derived from this analysis indicates that it is
convenient to take the value of the ratio equal to 0.2, within an 
accuracy of 10\%. 
The resulting $r$ estimate can then be used as the input value for a more 
refined modelling of the SED.

\begin{figure}
      \vspace{0.0cm}
      \hspace{0.0cm}
\epsfysize=10cm
\epsfbox{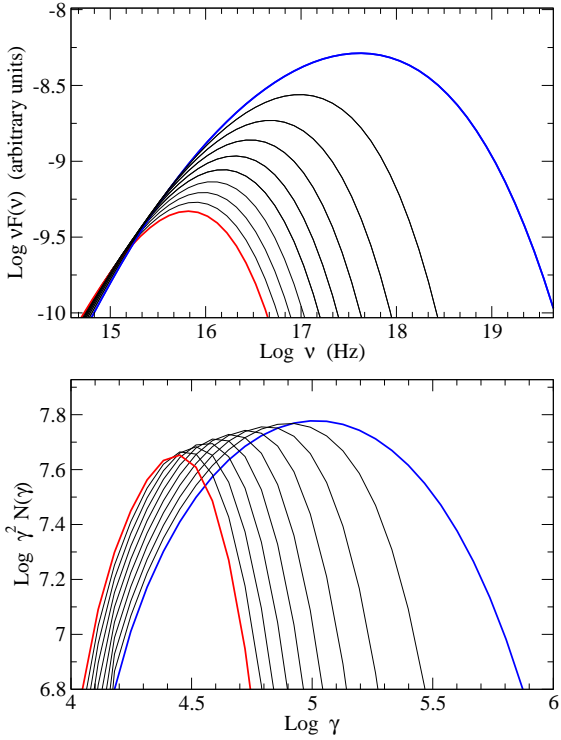}
\caption[]{
Time evolution of the SR spectrum (upper panel) and of the 
corresponding spectrum of electrons, multiplied by the square
of the Lorentz, under SR losses. 
The injection spectrum was a log-parabola with
$s=1.2$ and $r=0.7$, corresponding to a peak energy of $\sim 10^5$.
Spectra are plotted for equal time steps of $\sim$0.4 cooling times.
Note the decrease in the peak energy and the deviation 
from a log-parabola due to difference in cooling time with $\gamma$.
}
\label{fig4}
\end{figure}

\begin{figure}
      \vspace{0.0cm}
      \hspace{0.0cm}
\epsfysize=8cm
\epsfbox{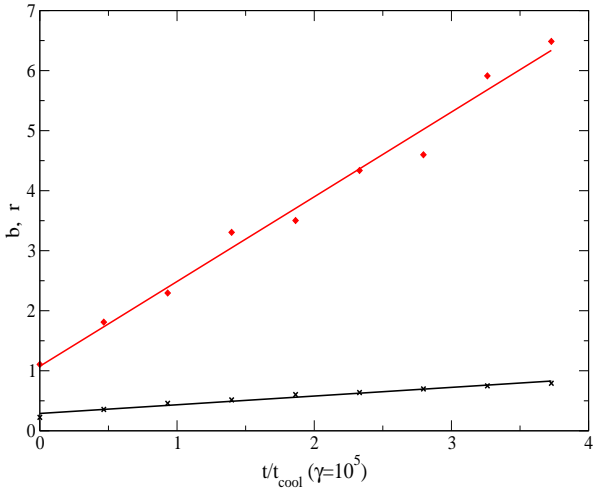}
\caption[]{
Time evolution of the log-parabolic curvature parameters 
$r$ of the electron spectrum (upper line) and $b$ of the SR SED 
(lower line) computed 
by means of a best fit around the peak. Solid lines are linear 
best fits to the computed values.
}
\label{fig5}
\end{figure}

\subsection{Spectral evolution under synchrotron cooling}

We also studied the evolution of spectral curvature due only to synchrotron
cooling. Of course, this study cannot be performed in general because
the solution of the time-dependent continuity equation is a function of
the assumed time evolution of the injection rate of emitting electrons
and of several other parameters like the mean residence time of particles
inside the emitting volume.
We studied, therefore, the simplest case of a pure radiative SR cooling
in a homogeneous and steady magnetic field without leakage and with a 
log-parabolic initial spectrum.
The time evolution of SR and electron spectra are shown 
in the two panels of Fig. 4, where the decrease in both the peak 
frequency and intensity with time is very evident.
Although radiation losses modify the spectral shape, it remains 
approximately log-parabolic, though with larger curvature parameters.
These variations are shown in Fig. 5, where we plotted the $r$ and $b$ 
values, measured in energy/frequency intervals centered approximately at 
the distribution peaks, as a function of time, measured in 
units of the cooling time for the peak energy of the initial electron
spectrum, taken as $\gamma$=10$^5$ and corresponding to $\sim$ 5$\times$10$^5$ s
with a magnetic field $B=0.1$ G.
It is worth noting that the change in time of $r$ and $b$ is described well 
by linear relations. The observation of a decrease in $\nu_{pS}$ 
associated with an increasing curvature during the dimming phase of 
an outburst is an indication of a radiative cooling evolution. 
Our calculations also show that the changes of SR peak frequency and $b$  
are correlated according to a power law as shown in the plot of Fig. 6.
 
We stress, however, that the estimate of $b$  may not be simple when 
working with observational data because the position of the peak 
may be far from the centre of the detector frequency range, and the 
uncertainty on the interstellar and intrinsic absorption may be 
non-negligible, particularly in the soft X-ray band.  

\begin{figure}
      \vspace{0.0cm}
      \hspace{0.0cm}
\epsfysize=8cm
\epsfbox{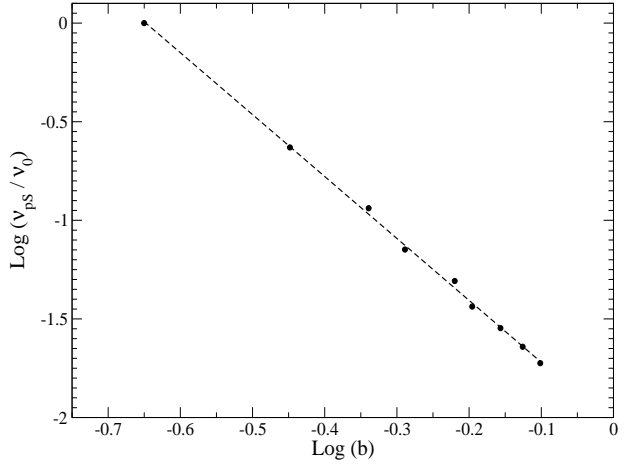}
\caption[]{
The relation between the peak frequency and the curvature parameter $b$ 
of the SR spectra plotted in the upper panel of Fig.~4. The dashed line is 
the best fit to the computed values.
}
\label{fig6}
\end{figure}

\subsection{Synchrotron Self-Compton spectrum}

We computed the spectral distribution of the SSC emission from
a relativistic electron population with a log-parabolic energy
spectrum in the framework of a homogeneous one-zone model in which
the seed photons are isotropic in the bulk frame of the
electrons. 
Our goal is to estimate the relations between the main spectral 
parameters: the peak frequencies of the SR and IC components and their 
curvatures indicated in the following by $b_S$ and $b_C$, respectively.

\begin{figure}
      \vspace{0.0cm}
      \hspace{0.0cm}
\epsfysize=10cm
\epsfbox{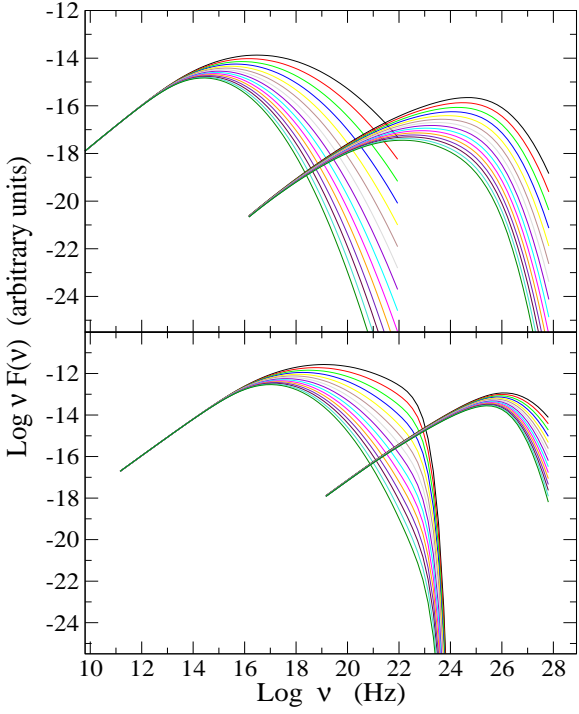}
\caption[]{
The SR and SSC spectral energy distributions emitted by an electron 
population with power-law log-parabolic distributions (Eq. 11), 
computed for $r$ values in the interval 0.50--1.20. 
Upper panel: spectra for $\gamma_0$=10$^3$; lower panel: spectra
for $\gamma_0$=2$\times$10$^4$.
}
\label{fig7}
\end{figure}

\begin{figure}
      \vspace{0.0cm}
      \hspace{0.8cm}
\epsfysize=10cm
\epsfbox{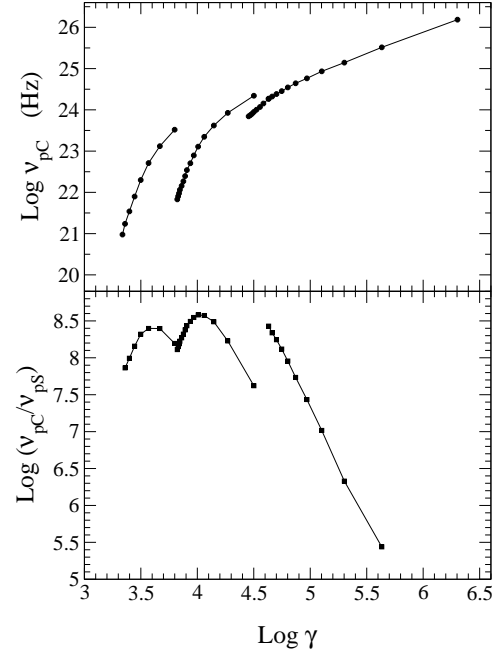}
\caption[]{
Dependence of the peak frequencies upon the particle characteristic energies 
in the SED of a single zone SSC model. 
Upper panel: frequency of SSC peak vs electron Lorentz factors of the 
peaks for $r$ values in the interval 0.50--1.90, the three curves
correspond to $\gamma_0$=10$^3$ (filled circles), 5$\times 10^3$ (open cicrcles), 
10$^4$ (crosses). 
Lower panel: ratio between the peak frequencies of IC and SR components
for the same values of the spectral parameters.
}
\label{fig8}
\end{figure}

The IC emissivity is given by (Jones 1968, Blumenthal \& Gould 1970, 
Band \& Grindlay 1985)
\begin{equation}
j_C(\nu')=(h \nu')~c \int d\nu~n(\nu) \int d\gamma~(d\sigma/d\nu')~N(\gamma) 
\end{equation}
where $n(\nu)$ is the density of SR photons, $\nu'$ the frequency of scattered 
photons, and the differential Klein-Nishina cross section is:
\begin{eqnarray}
\lefteqn{
\frac{d\sigma}{d\nu'} = \frac{3 \sigma_T}{16\gamma^4 \nu} \Big[ 2\eta~{\rm ln}\eta + 
      (1+2\eta)(1-\eta)+ }
       \nonumber \\
& & {}~~~~~~~~~~+ \frac{1}{2}(1-\eta)\frac{[4(h\nu/m_ec^2)\gamma \eta]^2}{1+4(h 
\nu/m_ec^2)\gamma\eta}\Big]{}
\end{eqnarray}
with 
\begin{equation}
0 \leq \eta = \frac{\nu'}{4 \gamma^2 \nu [1-(h \nu'/m_ec^2)]} \leq 1 
\end{equation}
and
\begin{equation}
1 \leq \frac{\nu'}{\nu} \leq \frac{4 \gamma^2}{1+ 4 \gamma(h \nu/m_ec^2)}  ~~~~~~. 
\end{equation}

Figure 7 shows two sets of resulting spectral distributions of both SR and IC 
components for values of $r$ ranging from 0.5 to 1.2 and $\gamma_0$ equal 
to 10$^3$ (upper panel) and 2$\times$10$^4$ (lower panel). 
The upper cut-off of the energy electron spectrum was fixed at 
$\gamma_{ct}$=5$\times$10$^7$.
In the upper panel of Fig. 8, we plotted the peak frequency of the IC 
component $\nu_{pC}$ vs $\gamma_p$ for $\gamma_0$=10$^3$, 5$\times 10^3$, 
10$^4$ and $r$ ranging from 0.5 to 1.9. 
These curves are useful for understanding at which electron peak energy the 
IC emission is dominated by the Thomson or the Klein-Nishina regime.
We stress that this does not imply that the largest contribution to 
the emission around $\nu_{pC}$ is due to electrons exactly at $\gamma_p$. 
We computed these energies and found that, for the highest value of 
$\gamma_0$, the difference is on the order of a few percent, while for the 
lowest $\gamma_0$ and $r$=1.7 the greatest contribution to the IC peak 
is produced by electrons having $\gamma \simeq$ 6.3$\times$10$^3$, 
while $\gamma_p$ is at 1.4$\times$10$^4$.   
It is of course impossible to derive a simple rule because of the number of 
parameters involved.
Figure 8 (upper panel) shows that as $\gamma_p$ increases, the corresponding 
$\nu_{pC}$ grows slower and slower indicating that the Klein-Nishina suppression 
becomes more efficient.
A second way to show the relation between SR and IC components is given in the 
lower panel of Fig. 8, where we plotted the ratio $\nu_{pC}/\nu_{pS}$ vs 
$\gamma_p$: this ratio increases until the Thomson scattering is dominant and 
reaches the maximum at the transition to the Klein-Nishina regime.

The IC spectra show an evident curvature that is not univocally related 
to the $r$ as in the SR case. Spectral curvature, in fact, depends on several 
parameters in a complex way.
The spectral shape is generally different from a log-parabola, and it can be 
approximated by this function only in a limited frequency interval. 
Consequently, the estimate of the curvature parameter
depends on the postion and amplitude of this interval.
In particular, the curvature depends on the intrinsic electron spectral curvature, 
the energy of SR photons, and on the energy of the electrons that mostly contribute 
to the emission in the selected interval. 
These energies, in fact, determine if the scatterings happen in the Thomson limit 
or in the Klein-Nishina regime. 
We obtained spectra with a curvature less pronounced than  
SR only for $\gamma_0$=10$^3$ (i.e. for dominant Thomson scattering) and $b$ 
evaluated in an interval around the peak. 
In all the other cases, higher $b$ values of IC spectrum resulted.
Figure 9 shows some examples: the curvature parameter 
$b$ of the IC spectrum, computed in a single zone SSC model, was evaluated in 
three adjacent frequency intervals, each having an amplitude of about
a decade starting from the peak frequency. 
When the main contribution to IC emission comes from Thomson scatterings, the curvature
is closer to that of SR (solid line), and it approaches that of the electron spectrum 
(dashed-dotted line) when the fraction of interactions in the Klein-Nishina regime
increases. 
The two panels correspond to $\gamma_0$ values differing by a factor of 20.
In the upper panel, IC emission is dominated by Thomson scattering in the two first
intervals, while in the lower panel the curvature approaches that of the electrons
because the majority of interactions are in the Klein-Nishina regime. 
This property is really useful because a simultaneous measure of the 
curvature parameters of SR and IC emissions can help to
discriminate between the two regimes of Compton scattering and to constrain 
the spectral parameters of emitting electrons.

\begin{figure}
      \vspace{0.0cm}
      \hspace{0.8cm}
\epsfysize=10cm
\epsfbox{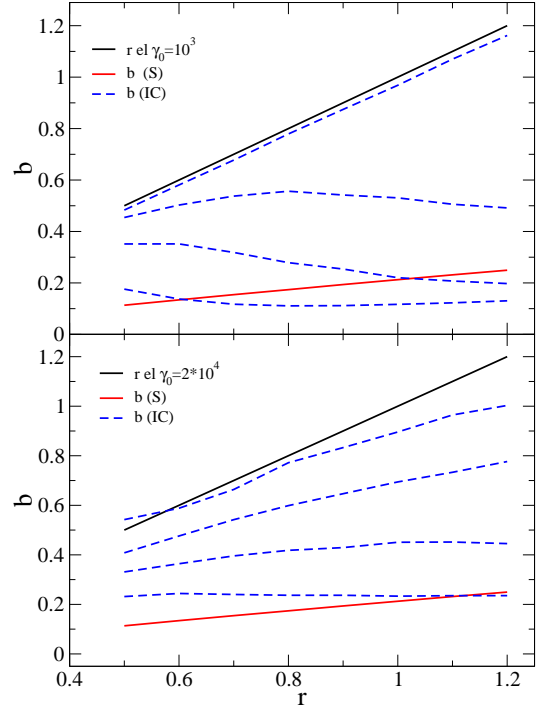}
\caption[]{
The curvature parameter $b$ of the IC spectrum for a single zone SSC model
plotted against the electron spectral curvature $r$. 
Dashed-dotted lines are the curvature parameters for the electrons and solid lines
for SR component. 
Dashed lines correspond to IC curvatures evaluated in three adjacent frequency
intervals with an amplitude of about a decade selected starting from $\nu_{pC}$:
lowest lines correspond to the intervals closest to the peaks and the highest ones
to those at the highest frequencies.  
The upper panel shows $b$ values computed for $\gamma_0$=10$^3$ and  lower panel those
for $\gamma_0$=2$\times$10$^4$.
Note that when the largest contribution to IC emission is from Thomson scatterings
(upper panel), curvatures are closer to those of SR, while in the Klein-Nishina regime
(lower panel) $b$ values result very close to $r$.
}
\label{fig9}
\end{figure}

\section{The X-ray and TeV emission from Mkn~501}
In Paper II we presented the results of analysing all BeppoSAX 
observations of the nearby HBL object Mkn~501 using log-parabolic 
spectral models.
In the case of this source, we did not detect significant variations 
of the X-ray flux during the individual pointings, but large luminosity 
changes were found between the three observations of April 1997 
(on the 7, 11, and 16) for which simultaneous TeV data, obtained 
with the CAT Cerenkov telescope, are available (Djannati-Atai et al. 1999). 
The TeV data showed a remarkable spectral curvature modelled by these 
authors by means of a log-parabolic law: best fit $b$ values were higher 
than 0.4, while those measured in the X rays are in the range 0.12--0.17 
(see Paper II). 
Krennich et al. (1999) observed Mkn~501 from February to June 1997 
with the Whipple telescope and found a high curvature of the TeV spectra 
as well.

In Sect. 4 we discussed the relations between the parameters of 
log-parabolic spectra for a SSC emission and showed that SSC emission 
model can give high frequency spectra with a $b$ value ranging from
that of the SR component to that of the electron spectrum depending on 
whether the scattering occurs in the Thomson or in the Klein-Nishina regime.
From the calculations of Sect. 3, we know that an SR $b$ value of about 0.15 
corresponds to $r\simeq$0.75, and therefore a curvature of the IC component
as that measured in the TeV range can be expected.

\begin{figure*}
  \vspace{1.0cm}
  \hspace{0.0cm}
  \epsfysize=11cm
  \epsfbox{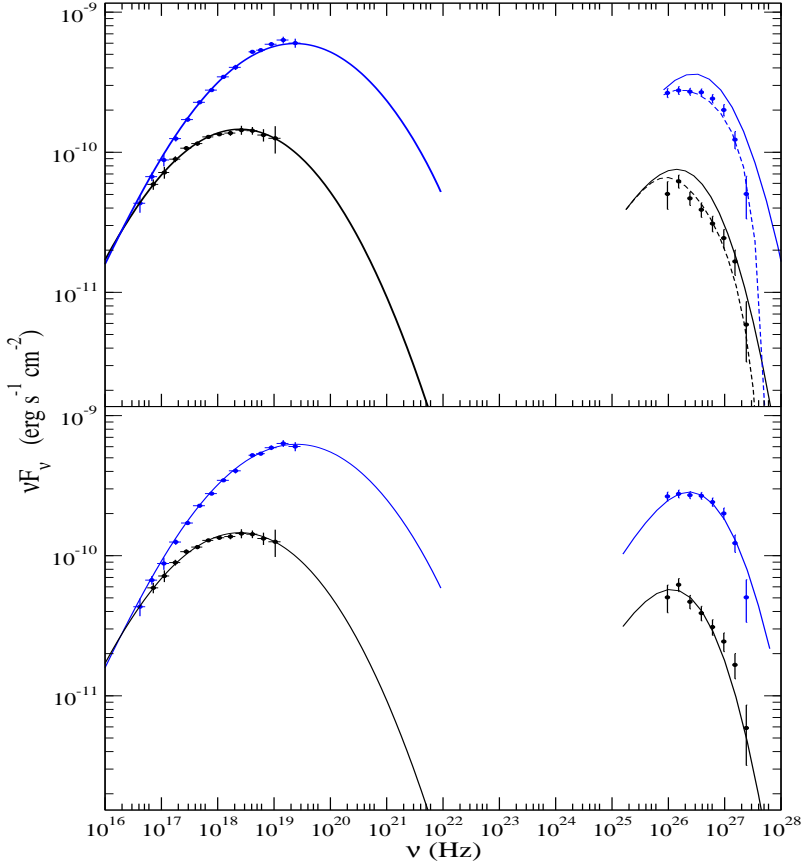}
  \caption[]{
    Two spectral energy distributions of Mkn~501 during the low and high states 
    observed  on 7 and 16 April 1997, respectively. X-ray points are from Paper II, 
    TeV points are  simultaneous CAT data (Djannati-Atai et al. 1999) and soolid lines
    are the spectra computed in a 1-zone SSC model for the SR and IC components.
    In the upper panel IC, spectra have been absorbed (dashed lines) by interaction with
    infrared EBL photons according to the LLL model by Dwek and Krennich (2005). 
    In the lower panel EBL absorption was neglected.
    }
\end{figure*}

In Paper II we also presented a two component-model to describe the 
spectral evolution of the April 1997 X-ray outburst of Mkn~501. 
We assumed that the components had the same curvature ($b=$0.18) and differ 
for the peak energy and their relative intensity. 
In this way we explained the nearly constant flux observed at lower energies
than $\sim$1 keV, while that around 100 keV changed by more 
than one order of magnitude, and the slight decrease of $b$ in the 
higher states was interpreted as the effect of the sum of two 
log-parabolic distributions peaked at different energies.

\begin{figure*}
  \vspace{1.0cm}
  \hspace{0.0cm}
  \epsfysize=11cm
  \epsfbox{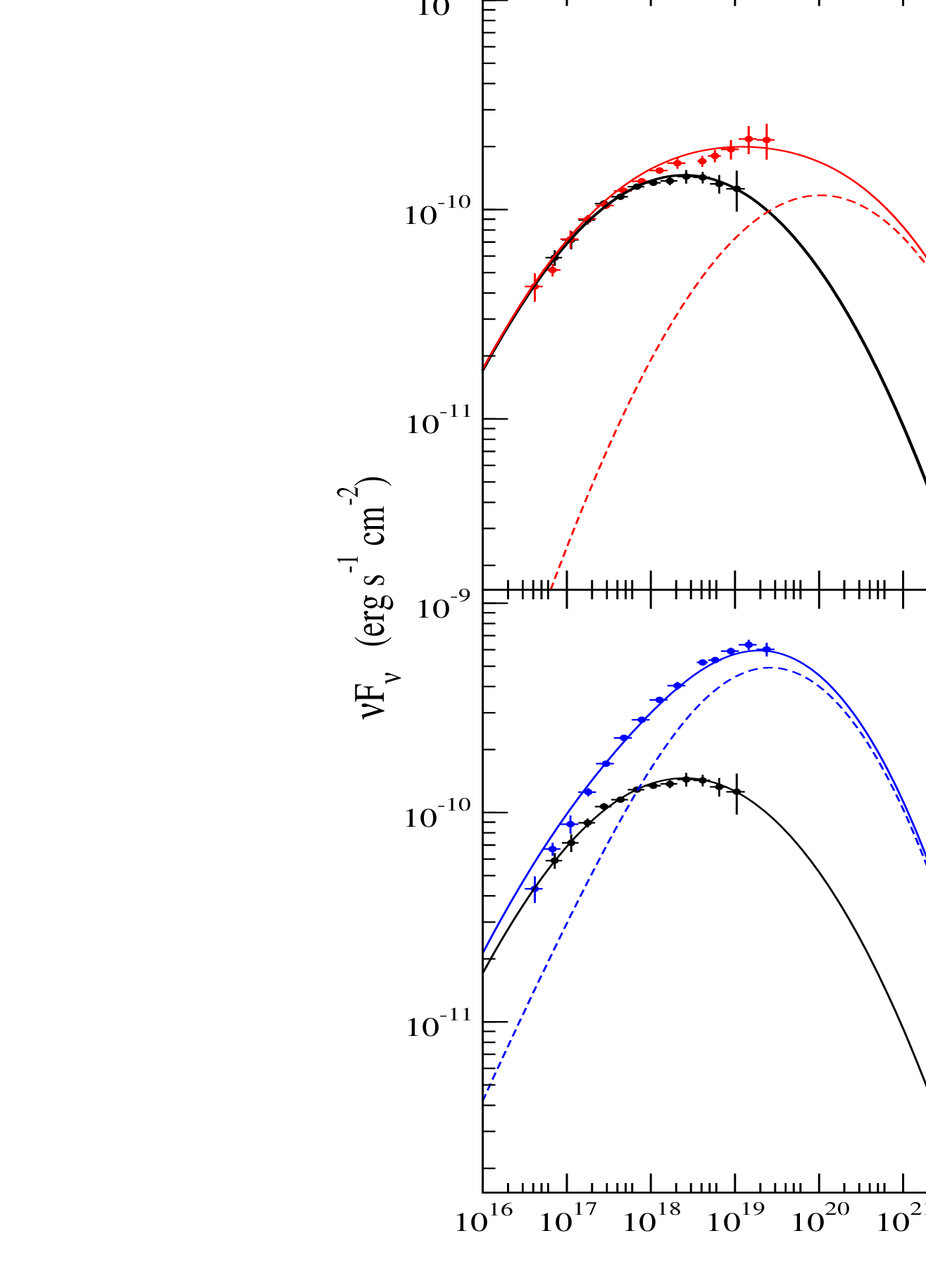}
  \caption[]{
    Two spectral energy distributions of Mkn~501 during the high states 
    observed  on 7, 11 April 1997 (upper panel) and 7, 16 April 1997 (lower panel). 
    X-ray points are from Paper II, 
    and TeV points are  simultaneous CAT data (Djannati-Atai et al. 1999). 
    Thin solid lines are the spectra computed in a 2-zone SSC model 
    for the SR and IC components, dashed lines are the spectra of the high-energy 
    flaring component, and the thick solid line is that of a slowly evolving component.
    }
\end{figure*}

An important effect to be considered when computing the spectral 
curvature in the TeV range is the absorption of the most energetic
photons due to pair production against the infrared extragalactic background
light (EBL). Dwek and Krennich (2005) have recently studied twelve
EBL models and computed the corresponding intrinsic spectra  
of Mkn~501. These authors found that the majority of EBL models give 
intrinsic emitted spectra showing an unphysical  rise at high 
TeV energies and that only the lowest intensity EBL models are acceptable.

We applied our radiation code to searching for the geometrical and 
physical parameters of the emitting region that can reproduce the observed 
intensities and spectral shapes, also taking EBL absorption into account.
To compare the model with the data, we need to introduce a relativistic
beaming factor $\delta$, which does not affect the spectral shape, 
but which changes observed frequencies and fluxes by $\delta/(1+z)$ and 
$[\delta/(1+z)]^3$, respectively, where $z=0.034$ is the source redshift. 
In the following we assume $\delta=15$, in agreement with other models 
for Mkn~501 (see e.g. Kataoka et al. 1999).

We  first considered a single-zone SSC model and tried to reproduce the 
two extreme states of the large outburst of April 1997, i.e. 
the low state on April 7 and the high state on April 16.

The spectrum of the emitting electrons was chosen with a curvature 
parameter $r$=$b$/0.2$\simeq 0.75$ following the relation found in Sect. 4.1, 
and the parameter $s$ was fixed to 1.2. 
Note that this value of $r$ is  consistent with the general behaviour 
presented in Fig. 9. 
The TeV curvature measured around the peak was about 0.4 for the low state 
and  0.45 for the high state (Djannati-Atai et al. 1999) and, for 
$\gamma_0$ on the order of $10^4$ and even higher, we expect a value of
$r$ in the range 0.5--0.8.
The values of the other parameters, given in Table 1, were chosen 
to achieve a satisfactory agreement between the data and the model.
It is interesting to note the behaviour of the two correlated parameters 
$\nu_{pS}$ (the peak frequency in the synchrotron SED) and 
$\gamma_p$ (the value of $\gamma$ of the electrons that contribute 
mostly to synchrotron peak emission). 
In fact the $\nu_{pS}$ grows during the flare by a factor of 11 
compared to the initial value while, according to Eq. (14), $\gamma_p$ 
changes as the square root of $\nu_{pS}$ growing rate.
Spectral distributions of Fig. 10 show that the SR follows the X-ray data 
accurately in both the spectral shape and peak 
evolution. The TeV spectra in the two states were computed using the SSC
code described in Sect. 4.2: in the upper panel of Fig. 10 we considered
the EBL absorption for the LLL model by Dwek and Krennich (2005),
while in the lower panel this absorption was neglected. In both cases
TeV data were well fitted in both the peak position and spectral curvature.
This suggests that this curvature could be partly intrinsic rather than
produced entirely by EBL absorption. The LLL model could then be
considered an upper limit to the extragalactic background.

\begin{table*}
\caption{ Model parameters for the 1997 April flare of Mkn~501.}
\label{tab1}
\begin{tabular}{lccccccccc@{ }cc}
\hline
Date & $s$ & $r$ & $N$ & $\gamma_0$ & $\gamma_p$ & $ V $ &
$E_e$ & $\nu L_\nu (SR)$ & $\nu L_\nu (IC)$ & $B$  & $\delta$  \\
     &     &     & cm$^{-3}$ & 10$^4$& 10$^5$& cm$^3$&  10$^{44}$ erg &10$^{39}$erg/s &
10$^{39}$erg/s & G  &   \\
\hline
SSC - 1 zone  &   &     &      &    &  &    &    &     &    \\
Apr. 07 & 1.12 & 0.81 & 4.25 & 4.80 &1.67&
$1.57\times$10$^{47}$& 91.0 & 8.70&4.47& 0.10 &15  \\
Apr. 16 & 1.18 & 0.70 & 30.0 & 8.70 &3.36   &
$1.72\times$10$^{46}$& 90.6 & 34.60 & 21.4 &0.14&15  \\ 

\hline
SSC - 2 zones  &   &     &   &    &    &     &    &     & & &    \\
Apr. 07 (Z1) & 1.14 & 0.81 & 4.6 & 4.0 & 1.35 & 1.2$\times$10$^{47}$ & 57.9 & 8.71 & 3.42 & 0.14 & 15  \\
Apr. 11 (Z2) & 1.2 & 1.00 & 1050 & 14 &3.52&9.2$\times$10$^{40}$ & 0.156 & 6.92& 1.44&1.5&15\\
Apr. 16 (Z2) & 1.2 & 1.33 & 250  & 20 &3.95&9.2$\times$10$^{43}$& 6.12 & 29.5& 14.1 &0.2&15\\ 
\hline
\multicolumn{10}{c} { }
\end{tabular}

$V$ is the volume of the emitting region; $E_e$ is the total energy of radiating electrons;
$\nu L_\nu (SR, IC)$ are the SED peak values of SR and IC components, respectively.

\end{table*}

As proposed in Paper II we considered a two-zone SSC model. 
A first zone (Z1) is responsible for the ``slowly" variable emission, 
whereas the second zone (Z2) is the source of the high-energy flare: 
the observed fluxes are then the sum of these two contributions. We used 
the April 7 emission as representative of the ``slowly'' variable state (Z1) and 
added the flaring component (Z2) to match the April 11 and 16 high states.
In this way the stability of the flux at energies $\leq$1 keV is
easily explained by the fact that the emission from Z2 is negligible when compared 
with that of Z1.
The results of these calculations are represented in the SEDs of Fig. 11.
We met some difficulty in obtaining a solution capable of modelling
the observed spectral curvature mostly in the highest state.
In fact, to deplete the SR from Z2 below $\sim$1 keV,
we need an electron population with a very low emissivity in
this band. This can be obtained only by introducing a low energy
cut-off, with the consequence of a high curvature of the low
energy portion of IC spectra at TeV energies.
The values of parameters for this 2-zone SSC model are also given 
in Table 1.

These results can be used to obtain some information on the
acceleration process. According to Eq.(10) we can derive the
energy gain as a function of $r$:
\begin{equation}
\varepsilon = 10^{q/2r}
\end{equation}
which, given that $r \simeq 1.0$, gives $\varepsilon \simeq 3.2^q$.
This implies that, unless $q$ is very small, the energy gain
is about a factor 3. Consequently, the
number of steps to increase the energy up to $\gamma \simeq 10^5$
is around 10. 
The acceleration must therefore be efficient and should probably 
occur for rather short times. 

It is interesting to compare our SEDs of the 1997 flare of Mkn~501 with 
those calculated by other authors who adopted different physical
models.
Generally, there is no explicit evaluation of a spectral curvature
parameter for both SR and IC components and consequently the electron
energy distributions are assumed to be power laws with exponential 
cut-off (Konopelko et al. 2003, Krawczynski 2002) or a broken power 
law (Katarzynski 2001).

To reproduce the spectral curvature in IC spectra, these authors  
need to introduce a heavy contribution from EBL interaction with 
TeV photons.
We stress that in our analysis the SR curvature is intrinsic, and we 
do not need break or cut-off to reproduce it. More interestingly, 
the IC curvature is also intrinsic and can be reproduced both using 
and neglecting the TeV photon absorption by EBL photons.
In particular in the, in case of large flares where the peaks of 
SR and IC components increase their energy, models with break or cut-off 
do not predict intrinsically curved IC spectra, and so the spectral shape 
observed at TeV and the IC peak frequency are mainly modulated by what 
EBL model is chosen. 
Note also that EBL attenuation could consistently modulate the position 
of the IC peak because its optical depth varies with the energy and
redshift. 
Observations of SR and IC components in objects with different redshifts 
and the study of the variations of their peak frequencies would be very
useful for more accurate modelling of the EBL spectrum.

It is interesting to compare the values of the main model parameters.
As a further point Konopelko et al. (2003) and Krawczynski et al. (2002) 
used $\delta$ values of 50 and 45, respectively, while Katarzynski et al. 
(2001) used 14, very close to our choice. 

\section{Summary and conclusion}
Over the past decade, broad-band observations of blazars have clearly shown that their
non-thermal SEDs are characterised by mildly curved shapes that can be
approximated by simple power laws only in frequency intervals that generally
not wider than a single decade. 
Starting from the first paper by Landau et al. (1986), we verified that 
a log-parabolic law is instead able to represent blazars' SEDs over several
frequency decades. 
An advantage of this law is that it has only three free parameters.
Furthermore, we have shown that for statistical acceleration mechanisms in which 
the acceleration probability is a decreasing function of the particle
energy, as in Eq. (7), the resulting spectrum follows a log-parabolic law.
Spectral properties of the SR can be derived easily only by using
the $\delta$-function approximation.

To find the exact relations between the spectral parameters of the energy
distribution of the relativistic electrons and of their SR, we 
used a precise numerical code specifically developed for blazar
applications. The main results can be summarised as follows:
\begin{itemize}
\item
The SR spectrum is log-parabolic with a curvature parameter approximately equal
to the electron curvature parameter multiplied by 0.2;
\item
spectral evolution due to SR cooling maintains a log-parabolic
distribution whose curvature parameter increases linearly with time;
\item
the IC SED around the peak can be approximately described by a log-parabola 
having a curvature parameter that ranges from values lower than the SR spectrum up to 
that of the electron distribution, depending upon whether the photon scattering is 
in the Thomson or Klein-Nishina regime.
The difference between the SR and IC curvatures can be used to 
discriminate between the two regimes.
\end{itemize}

We applied our code to the calculation of the SR and IC spectra of Mkn~501
when it was simultaneously observed by {\it BeppoSAX} (Paper II) and CAT (Djannati-Atai 
et al. 1999) during the large outburst of April 1997. We estimated the 
spectral parameters of the emitting electrons from X-ray data and then 
computed the IC spectrum in the TeV band, taking the absorption 
from pair production against EBL into account. 
Our results showed that the curvature at high energy is mostly 
intrinsic and larger than in the X-ray range, which shows that the IC 
scattering is taking place in the Klein-Nishina regime. 
This finding suggests that intergalactic absorption of TeV photons may 
be small, as shown by Dwek and Krennich (2005):
their LLL spectrum could be a satisfactory description of EBL, but we
cannot exclude that it has an even lower intensity.
This interpretation agrees with a previous work by
Krawczynski et al. (2000), where the low intensity DEBRA model 
(Malkan \& Stecker 1998) was favored over another model of higher intensity.
It is also supported by a recent paper by  Aharonian et al. (2005a), who
used H.E.S.S. data to find that in the time-averaged observations of Mkn 421,  
the cutoff energy of the IC spectrum is at $\sim$3.1 TeV, lower than that found
in Mkn 501 (about 6.2 TeV).
Considering the nearly equal redshifts of the these 
two blazars, they conclude that the cutoff is not due to EBL attenuation but 
is intrinsic to the sources, in agreement with our conclusion about 
the IC curvature.
We conclude (i) that the extragalactic space is probably more 
transparent to TeV photons than previously assumed and (ii) that a larger
number of blazars at higher redshifts than previously thought
could be detected in this range.

The same conclusion has been reached very recently by Aharonian et
al. (2005b), who report the detection  of 
the blazar 1ES 1101$-$232 at $z = 0.186$ (by H.E.S.S. in the TeV band).
An important consequence of the intrinsic curvature of TeV
spectra is that it can be variable in time and possibly related to the X-ray
curvature.
As a final remark, we stress the relevance of the simultaneous broad-band 
observations that are necessary to obtain a reliable estimate of
curvature parameters for both components.

\begin{acknowledgements}
We acknowledge the helpful comments and suggestions of the referee, 
A. Djannati-Atai. This work was financially supported for the
ASDC by the Italian Space Agency (ASI) and for the Physics
Dept. by Universit\`a di Roma La Sapienza. 

\end{acknowledgements}


\begin{thebibliography}{ }
\small
\bibitem[Aharonian 2005a]{aharoa}
Aharonian F., Akhperjanian A.G. et al.,
2005, A\&A 437, 95

\bibitem[Aharonian 2005b]{aharob}
Aharonian F., Akhperjanian A.G. et al.,
2005, Nature, submitted, astro-ph/0508073 

\bibitem[Band 1993]{band} 
Band D. et al., 
1993, ApJ 413, 281

\bibitem[Bell 1978]{bell} 
Bell A.R., 
1978, MNRAS 182, 147

\bibitem[Blandost 1979]{blost} 
Blanford R.D., Ostriker J.P., 
1978, ApJ 221, L29

\bibitem[Blumgould 1970]{blugou} 
Blumenthal G.R., Gould R.J., 
1970, Rev. Mod. Phys. 42, 237

\bibitem[Djannati 1999]{djanna} 
Djannati-Atai A., Piron F., Barrau A. et al., 
1999, A\&A 350, 17

\bibitem[Dwekrennich 2005]{dwekren} 
Dwek E., Krennich F.,  
2005, ApJ 618, 657

\bibitem[Kardashev 1962]{karda} 
Kardashev N., 
1962, Soviet Astron. 6, 317

\bibitem[Kataoka 1999]{kataoka} 
Kataoka J., Mattox J.R. et al.,
1999, ApJ 514, 138

\bibitem[Katarzynski 2001]{katar} 
Katarzy\~nski K., Sol H., Kus A., 
2001, A\&A 367, 809

\bibitem[Konopelko 2003]{konop} 
Konopelko A., Mastichiadis A. et al.,
2003, ApJ 597, 851

\bibitem[Krawczynski 2000]{kraw00}
Krawczynski, H., Coppi, P.S., Maccarone, T., Aharonian, F.A.
2000, A\&A 353, 97

\bibitem[Krawczynski 2002]{kraw} 
Krawczynski H., Coppi P.S., Aharonian F., 
2002, MNRAS 336, 721

\bibitem[Krennich 1999]{krenni} 
Krennich F., Biller S.D., Bond I.H. et al., 
1999, ApJ 511, 149

\bibitem[Landau 1986]{landau} 
Landau R. et al., 
1986, ApJ 308, 78

\bibitem[Malkan 1998]{Malkan}
Malkan M.A., Stecker F.W., 
1998, ApJ 496, 13

\bibitem[Marcher et al. 1985]{march85} 
Marscher A.P., Gear W. K.,
1985, ApJ 298, 114

\bibitem[Massaro et al. 2004]{massaro} 
Massaro E., Perri M., Giommi P., Nesci R., 
2004a, A\&A 413, 489 (Paper I)

\bibitem[Massaro et al. 2004]{mass04} 
Massaro E., Perri M., Giommi P. et al., 
2004b, A\&A 422, 103 (Paper II)

\bibitem[Michel 1981]{michel} 
Michel F.C., 
1981, ApJ 247, 664

\bibitem[Padovani 1995]{padov} 
Padovani P., Giommi P., 
1995, ApJ 111, 222

\bibitem[Nodes et al. 2004]{nodes} 
Nodes C., Birk G.T., Gritschneder M., Lesch H., 
2004, A\&A 423, 13 

\bibitem[Perri 2003]{perri} 
Perri M., Massaro E., Giommi P. et al., 
2003, A\&A 407, 453 

\bibitem[Perlman 2005]{perl5} 
Perlman E.S., Madejski G. et al., 
2005, ApJ (in press) astro-ph/0502298 

\bibitem[Piranomonte et al. 2002]{Pir02} 
Piranomonte S., Giommi P., Verrecchia F., Perri M., 
2002, in ``Blazar Astrophysics with BeppoSAX and other Observatories", eds.
P. Giommi, E. Massaro \& G. Palumbo, ASI Special Publ., p. 51

\bibitem[Sambruna 1996]{sambr96}
Sambruna R., Maraschi L., Urry M.C.,
1996, ApJ 463, 444

\bibitem[Sikora 1994]{sikora94}
Sikora M., Begelman M.C., Rees M.J.,
1994, ApJ 421, 123

\bibitem[Sikora 2002]{sikora}
Sikora M., Blazejowski M. et al.,
2002, ApJ 577, 78

\bibitem[Sohn 2003]{sohn} 
Sohn B.W., Klein U., Mack K.-H., 
2003, A\&A 404, 133 

\bibitem[Tavecchio 2001]{tavec01} 
Tavecchio F.,  Maraschi L., Pian E. et al., 
2001, ApJ 554, 725

\bibitem[Tramacere 2002]{tram02} 
Tramacere A., 2002, Laurea thesis, Univ. of Perugia 

\bibitem[Tramacere 2003]{tram03} 
Tramacere A., Tosti G., 2003, New AR 47, 697 

\end{thebibliography}
\end{document}